\documentclass[12pt]{emulateapj}

\usepackage{natbib}

\newcommand{\teff}{${T}_{\mathrm{eff}}$}
\newcommand{\logg}{$\log{g}$}
\newcommand{\msun}{$M_{\odot}$}
\newcommand{\rsun}{$R_{\odot}$}
\newcommand{\kms}{km s$^{-1}$}
\newcommand{\muhz}{$\mu$Hz}

\shorttitle{The First Pulsating Extremely Low Mass WD}
\shortauthors{Hermes et al.}

\begin{document}

\title{SDSS J184037.78+642312.3: THE FIRST PULSATING EXTREMELY LOW MASS WHITE DWARF}
\author{J. J. Hermes\altaffilmark{1,2}, M. H. Montgomery\altaffilmark{1,2}, D. E. Winget\altaffilmark{1,2}, Warren R. Brown\altaffilmark{3}, \\ Mukremin Kilic\altaffilmark{4} and Scott J. Kenyon\altaffilmark{3}}

\altaffiltext{1}{Department of Astronomy, University of Texas at Austin, Austin, TX\,-\,78712, USA}
\altaffiltext{2}{McDonald Observatory, Fort Davis, TX\,-\,79734, USA}
\altaffiltext{3}{Smithsonian Astrophysical Observatory, 60 Garden St, Cambridge, MA\,-\,02138, USA}
\altaffiltext{4}{Homer L. Dodge Department of Physics and Astronomy, University of Oklahoma, 440 W. Brooks St., Norman, OK\,-\,73019, USA}

\email{jjhermes@astro.as.utexas.edu}

% ++++++++++++++++++++++++++++++++++++++++++++++++++++++++++++++++++++ %
% ++++++++++++++++++++++++++++++ Abstract ++++++++++++++++++++++++++++ %
% ++++++++++++++++++++++++++++++++++++++++++++++++++++++++++++++++++++ %

\begin{abstract}

We report the discovery of the first pulsating extremely low mass (ELM) white dwarf (WD), SDSS J184037.78+642312.3 (hereafter J1840). This DA (hydrogen-atmosphere) WD is by far the coolest and the lowest-mass pulsating WD, with \teff=$9100\pm170$ K and \logg=$6.22\pm0.06$, which corresponds to a mass $\sim 0.17$ \msun. This low-mass pulsating WD greatly extends the DAV (or ZZ Ceti) instability strip, effectively bridging the $\log g$ gap between WDs and main sequence stars. We detect high-amplitude variability in J1840 on timescales exceeding 4000 s, with a non-sinusoidal pulse shape. Our observations also suggest that the variability is multi-periodic. The star is in a 4.6 hr binary with another compact object, most likely another WD. Future, more extensive time-series photometry of this ELM WD offers the first opportunity to probe the interior of a low-mass, presumably He-core WD using the tools of asteroseismology.

\end{abstract}


% ++++++++++++++++++++++++++++++++++++++++++++++++++++++++++++++++++++ %
% +++++++++++++++++++++++++++++++ INTRO ++++++++++++++++++++++++++++++ %
% ++++++++++++++++++++++++++++++++++++++++++++++++++++++++++++++++++++ %

\section{Introduction}

Asteroseismology allows us to probe below the photosphere and into the interiors of stars. There are many pulsational instability strips on the Hertzsprung-Russell diagram, including the DAV (or ZZ Ceti) instability strip, driven by a hydrogen partial ionization zone in the hydrogen atmosphere (DA) WDs. Seismology using the non-radial $g$-mode pulsations of DAVs enables us to constrain the mass, core and envelope composition, rotation rate, and the behavior of convection in these objects (see reviews by \citealt{WinKep08} and \citealt{FontBrass08}).

The mass distribution of DA WDs in the SDSS shows a strong peak at 0.6 \msun with tails toward higher and lower masses \citep{Tremblay11}; masses of individual WDs range from about 0.2 \msun\ to 1.3 \msun. The roughly 150 DAVs known to date have masses $0.5-1.1$ \msun, implying they all likely contain C/O-cores. Lower mass WDs are likely to pulsate as well. However, previous searches have failed to detect such pulsations \citep{Steinfadt12}.

The galaxy is not old enough to produce low mass ($<0.5$ \msun) WDs through single-star evolution; these WDs are believed to be the product of binary evolution. Indeed, radial velocity surveys of low-mass WDs indicate that most form in binary systems \citep{Marsh95,Brown11a}. Many short-period binaries go through one or two common-envelope phases, which may effectively remove enough mass to prevent ignition of He to C/O. However, there is little direct evidence that low-mass WDs have He-cores. But if they pulsate, as do their C/O-core brethren, we may differentiate their interior structure.

We have been engaged in an ongoing search for low-mass DAVs for many years at McDonald Observatory. The benefits of a search for a low-mass (and putatively He-core) DAV were recently emphasized by \citet{Steinfadt10}. Should they pulsate in $g$-modes like the C/O-core DAVs, the eigenfunctions of ELM WDs would globally sample the interior, making the pulsations sensitive to core composition. Seismology may also allow us to constrain the hydrogen layer mass; this is vitally important since hydrogen burning is expected to be a major or even dominant component of the luminosity of these stars \citep[e.g.,][]{Panei07}.

\citet{Steinfadt12} outlined the null results of a search for pulsations in 12 low-mass WDs. We have extended a similar, systematic search for variable He-core WDs, armed with the many dozens of new extremely low-mass (ELM, $\sim0.2$ \msun) WDs catalogued by the ELM Survey \citep{BrownELMi,KilicELMii,BrownELMiii,KilicELMiv}. That search has yielded its first success.
 
In this Letter, we report the discovery of pulsations in the ELM WD SDSS J184037.78+642312.3, and show that the photometric variations are likely multi-periodic. We have also included our null results for another eight low-mass WDs that were observed not to vary, to various detection limits.

% ==================================================================== %
% ==================================================================== %

% ++++++++++++++++++++++++++++++++++++++++++++++++++++++++++++++++++++ %
% ++++++++++++++++++++++++++++ Observations ++++++++++++++++++++++++++ %
% ++++++++++++++++++++++++++++++++++++++++++++++++++++++++++++++++++++ %

\section{Observations}
\label{sec:obs}
% ++++++++++++++++++++++++++++ J1840 LC ++++++++++++++++++++++++++++++ %
\begin{figure*}[t]
\centering{\includegraphics[width=0.8\textwidth]{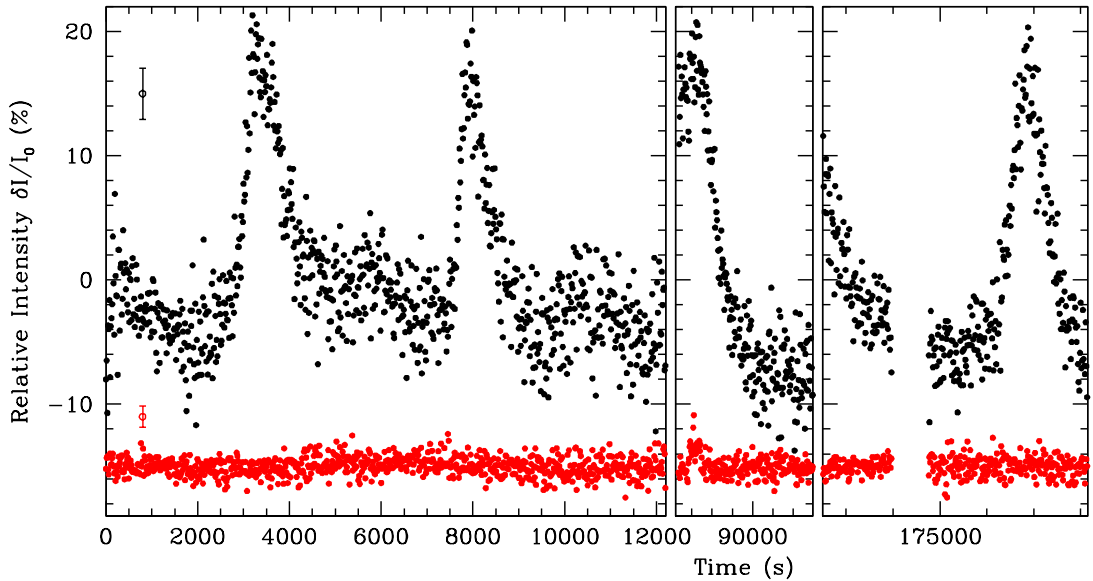}}
\caption{High-speed photometry of J1840 (in black) over three consecutive nights in 2011 October. The gap in the third night was caused by a passing cloud. The bottom red light curve, offset by $-15$\%, shows the brightest comparison star in the field over the same period. Average 2$\sigma$ errors are shown offset with error bars. \label{fig:J1840lc}}
\end{figure*}
% ==================================================================== %

\citet{BrownELMiii} present the spectroscopic discovery data for J1840 from the Blue Channel spectrograph on the 6.5m MMT. They use 37 separate spectra over more than a year to determine the system parameters. \citet{BrownELMiii} find that J1840 is in a $4.5912\pm0.0012$ hr ($16528.32\pm4.32$ s) orbital period binary with a $K=272\pm2$ \kms\ radial velocity semi-amplitude. However, a significant alias exists at 3.85 hr.

Model fits to the co-added spectra for J1840 yield \teff\ $=9140\pm170$ K and \logg\ $= 6.16\pm0.06$, which correspond to a mass of roughly 0.17 \msun. Given the mass function of the system, $f = 0.399\pm0.009$ \msun, the minimum mass of the unseen companion is 0.64 \msun; if the orbital inclination is random, there is a 70\% probability that the companion is a WD with $<1.4$ \msun. We note, however, that the nature of the unseen companion has no bearing on the impact of this Letter.

We obtained high speed photometric observations of J1840 at the McDonald Observatory over three consecutive nights in 2011 October, for a total of more than 5.5 hr of coverage. We used the Argos instrument, a frame-transfer CCD mounted at the prime focus of the 2.1m Otto Struve telescope \citep{Nather04}, to obtain many 15 s exposures on this $g_0=18.8$ mag WD. The seeing averaged 1.5\arcsec\ and transparency variations were low, although our second and third nights were cut short by clouds. Observations were obtained through a 1mm BG40 filter to reduce sky noise.

We performed weighted aperture photometry on the calibrated frames using the external IRAF package $\textit{ccd\_hsp}$ written by Antonio Kanaan (the reduction method is outlined in \citealt{Kanaan02}). We divided the sky-subtracted light curves by five brighter comparison stars in the field to allow for fluctuations in seeing and cloud cover, and applied a timing correction to each observation to account for the motion of the Earth around the barycenter of the solar system \citep{Stumpff80,Thompson09}.

Figure~\ref{fig:J1840lc} shows all $1365$ Argos light curve points obtained for J1840 from 25 Oct 2011 to 27 Oct 2011. We also include the light curve of the brightest comparison star in the field, SDSS J184043.21+642351.8, for reference.

% ==================================================================== %
% ==================================================================== %

% ++++++++++++++++++++++++++++++++++++++++++++++++++++++++++++++++++++ %
% +++++++++++++++++++++++++++++ Analysis +++++++++++++++++++++++++++++ %
% ++++++++++++++++++++++++++++++++++++++++++++++++++++++++++++++++++++ %

\section{Analysis}

Our photometric data set is relatively short, as we caught J1840 just before it went behind the Sun. This limits the significance of the detected periods, and we eagerly anticipate further observations. Still, we have sufficient data to show convincingly that this low-mass WD is a multi-periodic variable star.

% ++++++++++++++++++++++++++++ J1840 Fs ++++++++++++++++++++++++++++++ %
\begin{deluxetable}{cccc}
\tablecolumns{4}
\tablewidth{0.45\textwidth}
\tablecaption{Frequency solutions for SDSS J1840+6423
  \label{tab:freq}}
\tablehead{\colhead{Period} & \colhead{Frequency} & \colhead{Amplitude} & S/N 
\\ \colhead{(s)} & \colhead{($\mu$Hz)} & \colhead{(\%)} & \colhead{} }
\startdata
\multicolumn{4}{c}{\bf Multi-mode solution I} \\
\hline\\
4445.9 $\pm$ 1.4 & 224.926 $\pm$ 0.070 & 6.59 $\pm$ 0.19 & 8.0 \\
2376 $\pm$ 57 & 420 $\pm$ 10 & 4.883 $\pm$ 0.83 & 6.3 \\
1578.56 $\pm$ 0.37 & 633.49 $\pm$ 0.15 & 2.95 $\pm$ 0.25 & 4.4 \\
%Residuals: 0.03727
\hline\\
\multicolumn{4}{c}{\bf Multi-mode solution II} \\
\hline\\
4445.3 $\pm$ 2.4 & 224.96 $\pm$ 0.12 & 7.6 $\pm$ 1.6 & 9.4 \\
2376.07 $\pm$ 0.74 & 420.86 $\pm$ 0.13 & 4.817 $\pm$ 0.46 & 6.3 \\
1578.70 $\pm$ 0.65 & 633.43 $\pm$ 0.26 & 2.831 $\pm$ 0.41 & 4.3 \\
3930 $\pm$ 300 & 254 $\pm$ 19 & 2.7 $\pm$ 2.0 & 3.4 \\
1164.15 $\pm$ 0.38 & 859.00 $\pm$ 0.29 & 1.78 $\pm$ 0.29 & 3.3 \\
%Residuals: 0.03045
\hline\\
\multicolumn{4}{c}{\bf Single-mode harmonic solution} \\
\hline \\
4443.77 $\pm$ 0.80 & 225.034 $\pm$ 0.041 & 7.35 $\pm$ 0.19 & 5.7 \\
2221.89 $\pm$ 0.40 & 450.068 $\pm$ 0.081 & 4.03 $\pm$ 0.19 & 3.5 \\
1481.26 $\pm$ 0.27 & 675.10 $\pm$ 0.12 & 1.53 $\pm$ 0.18 & 1.6
%Residuals: 0.04561
\enddata
\end{deluxetable}
% ==================================================================== %

The high-amplitude variability is easy to distinguish in the raw light curve (Figure~\ref{fig:J1840lc}), with more than 25\% peak-to-peak variability. The highest peak in a Fourier transform (FT) of the brightest companion star in the field yields only a small signal (0.24\% amplitude) at $7340\pm15$~s, consistent with low-frequency noise from atmospheric variability. Our apertures and sky annuli have been chosen to ensure there is no significant contamination from the nearby star SDSS J184038.73+642315.6, which is 7.0\arcsec\ away from our target. Thus the signal we are observing is intrinsic to the WD. Without evidence for a companion star or accretion from the spectra, the light curves, or existing broadband photometry, we are confident that the photometric variability results from pulsations on the surface of the ELM WD.

The pulse shape appears non-sinusoidal, with a steep rise and decline. We first test whether a single mode ($f_1$) and its harmonics ($2f_1 + 3f_1 + ...$) could reproduce the observed light curve. A nonlinear least squares fit with the highest peak in the FT and its fixed harmonics converges on 225.03 \muhz\ (4444 s) as the best parent mode (see the bottom of Table~\ref{tab:freq} for a full solution).

However, our data from multiple nights rule out this scenario. The single-mode harmonic solution fits the two peaks during the first night well, but it fails to match the peak in our second night of data, predicting a maximum in the light curve more than 500 s too soon.

% ++++++++++++++++++++++++++++ J1840 FT ++++++++++++++++++++++++++++++ %
\begin{figure}[t]
\centering{\includegraphics[width=0.975\columnwidth]{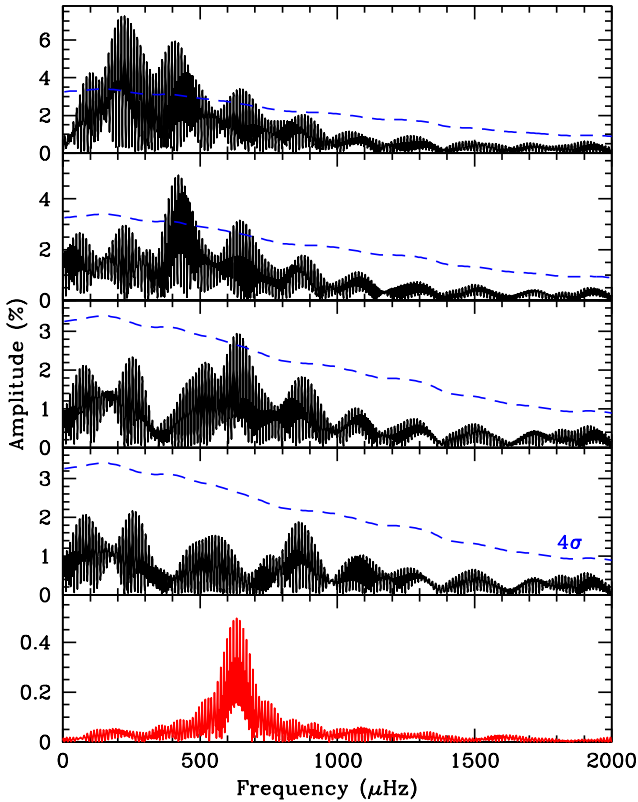}}
\caption{Fourier transforms of the light curve. The top panel shows the original data, the second panel the data after prewhitening by the peak at 224.926 \muhz\ (4445.9 s), the third panel after prewhitening by the peak at 420 \muhz\ (2376 s), and the fourth panel after prewhitening by the peak at 633.49 \muhz\ (1578.56 s). See Table~\ref{tab:freq} for our frequency solution. The dashed blue line shows 4 times the noise level, as described in the text. The bottom panel shows the spectral window folded around 633.49 \muhz. \label{fig:J1840ft}}
\end{figure}
% ==================================================================== %

A multi-mode solution (see the top panels of Table~\ref{tab:freq}) where the frequencies are determined from the highest amplitudes in an FT of the entire dataset (see Figure~\ref{fig:J1840ft}) improves the residuals by more than 20\%. For more realistic estimates, the cited errors are not formal least-squares errors to the data but rather the product of $10^5$ Monte Carlo simulations of perturbed data using the software package Period04 \citep{Lenz05}. The signal-to-noise calculation is based on the average amplitude of a 1000 \muhz\ box after pre-whitening by the three significant, highest-amplitude frequencies.

A small but possibly significant amount of power remains around 254 and 859 \muhz\ after prewhitening by the three significant periodicities (with S/N $> 4$), so we have included a frequency solution with these additional periods in Table~\ref{tab:freq}. This set of period solutions is by no means exhaustive, but it establishes that this ELM WD is variable and multi-periodic.

% ==================================================================== %
% ==================================================================== %

% ++++++++++++++++++++++++++++++++++++++++++++++++++++++++++++++++++++ %
% +++++++++++++++++++++++++++++ Discussion +++++++++++++++++++++++++++ %
% ++++++++++++++++++++++++++++++++++++++++++++++++++++++++++++++++++++ %

\section{Discussion}

We announce the discovery of the first pulsating extremely low mass white dwarf, J1840, which is both the coolest and lowest-mass WD known to pulsate. The object offers, for the first time, an opportunity to explore the interior of a putative He-core WD using asteroseismology. 

Asteroseismology of ELM WDs will help constrain the thickness of the surface hydrogen layers in these low-mass WDs. There are several millisecond pulsars with ELM WD companions, and the cooling ages of such companions can be used to calibrate the spin-down ages of these pulsars. However, current evolutionary models for ELM WDs are relatively unconstrained. For masses $M>0.17~M_\odot$, diffusion-induced hydrogen-shell flashes take place, which yield small hydrogen envelopes \citep{Althaus01,Panei07,Kilic10a}. The models with $M \leq 0.17~M_\odot$ do not experience thermonuclear flashes. As a result, they have massive hydrogen envelopes, larger radii, lower surface gravities, and they are predicted to evolve much more slowly compared to more massive WDs. If enough modes are excited to observability, we hope to directly constrain the hydrogen layer mass.

A plethora of excited modes would also allow for measuring the mean period spacing, which is a sensitive function of the mass of the star, and is also slightly dependent on the core mass fraction. The models of \citet{Steinfadt10} found a mean period spacing of $\sim 89$ s for $\ell=1$ $g$-modes of a 0.17 \msun\ ELM WD, about a factor of two larger than the observed $47\pm12$ s period spacing for $\ell=1$ $g$-modes in the cool C/O-core counterpart G29-38 \citep{Kleinman98}.

We do expect this non-radial pulsator to be multi-periodic even if just one pulsation mode and its non-linear harmonics are amplified to observability: J1840 is in a relatively close binary with an unseen compact object. This companion will influence the light curve in many ways. Although we lack the sensitivity and full phase coverage needed for a detection, we expect a $\sim0.3$\% amplitude Doppler beaming signal at the orbital period given the effective temperature of J1840 and the radial-velocity amplitude \citep{Shporer10}. The companion will also induce tidal distortions on the primary, as seen in many other ELM WDs \citep{KilicJ0106,BrownJ0651,Hermes12}. However, assuming that J1840 has a radius of roughly 0.054 \rsun\ \citep{Panei07}, ellipsoidal variations should be smaller than 0.1\% for even the highest possible inclinations \citep{Morris93}.

The companion's effect on the rotation period is perhaps more significant. The $\sim 1.7$ Gyr cooling age of this ELM WD \citep{Panei07} may be longer than the synchronization timescale for such a short-period binary \citep[e.g.,][]{Claret95}. If so, and the ELM WD is rotating at the orbital period of 4.6 hr, then non-radial pulsations will be subsequently subject to rotational splittings determined by their modal degree. For example, if the 225 \muhz\ mode is an $\ell = 1$ mode, a 4.6 hr rotation rate would cause it to be split by about 30 \muhz, assuming solid-body rotation \citep{Unno89}. However, synchronization is not a guarantee: Recent \emph{Kepler} observations of a close sdB+dM binary found the primary rotating much slower than the $\sim 9.5$ hr orbital period \citep{Pablo12}. An analysis of the rotational splittings in the pulsations of J1840 will test tidal synchronization in this system, thereby probing the rigidness of this ELM WD.

% ++++++++++++++++++++++++++++++ Search ++++++++++++++++++++++++++++++ %
\begin{figure}[t]
\centering{\includegraphics[width=\columnwidth]{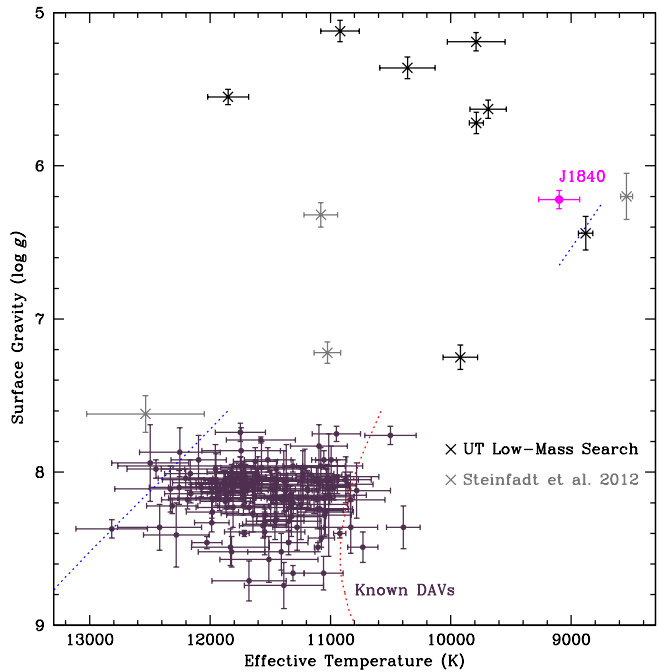}}
\caption{Search parameter space for pulsations in low-mass WDs. Known DAVs are included as purple dots, and our new pulsator is labeled and marked as a magenta dot. DA WDs observed not to vary to at least 1\% are marked with an X; those from \citet{Steinfadt12} are marked in gray and those from our search (see Table~\ref{tab:null}) in black. To guide the eye, we have marked the rough boundaries of the observed DAV instability strips as dotted lines; the short, blue dotted line marks the blue-edge of the theoretical \citet{Steinfadt10} instability strip. \label{fig:search}}
\end{figure}
% ==================================================================== %

We also hope to use the nonlinearities in this non-sinusoidal light curve to constrain the size of the convection zone of this WD \citep[e.g.,][]{MikeMon05,MikeMon10}. Since ELM WDs exist in a completely new regime of parameter space than do typical C/O core WDs, this analysis will provide important independent constraints on the structure of their outer layers, as well as providing a measure of the convective efficiency in this new regime.

We discovered the pulsations in J1840 as part of a systematic search for pulsations in low-mass WDs, recently energized by the work of \citet{Steinfadt10, Steinfadt12}. We have explored a vast area of the \logg\ --- \teff\ parameter space for variability (see Figure~\ref{fig:search}), and compiled a list of our first eight null results in Table~\ref{tab:null}. These null observations were reduced and analyzed in an identical manner to those outlined in Section~\ref{sec:obs}, and the atmospheric parameters and their formal errors were determined by the references cited.

% ++++++++++++++++++++++++++++++ Table +++++++++++++++++++++++++++++++ %
\begin{deluxetable*}{lcccrc}
\tablecolumns{6}
\tabletypesize{\footnotesize}
\tablewidth{1.0\textwidth}
\tablecaption{Observed Low-Mass DAV Candidates and Null Results\label{tab:null}}
\tablehead{\colhead{Object} & \colhead{$g^\prime$-SDSS} & \colhead{\teff} &
 \colhead{\logg} & \colhead{Reference} & \colhead{Det. Limit} \\ 
 \colhead{} & \colhead{(mag)} & \colhead{(K)} & \colhead{(cm s$^{-1}$)} & \colhead{} & \colhead{\%} }
\startdata
SDSS~J011210.25+183503.74 & 17.3 & $9690\pm150$  & $5.63\pm0.06$ & \citet{BrownELMiii} & 0.1 \\
SDSS~J082212.57+275307.4  & 18.3 & $8880\pm\phantom{1}60$   & $6.44\pm0.11$ & \citet{Kilic10a} & 0.2 \\
SDSS~J091709.55+463821.8  & 18.7 & $11850\pm170\phantom{1}$ & $5.55\pm0.05$ & \citet{Kilic07} & 0.3 \\
SDSS~J122822.84+542752.92 & 19.6 & $9921\pm143$  & $7.25\pm0.08$ & \citet{Eisenstein06} & 0.7 \\
SDSS~J123316.20+160204.6  & 19.8 & $10920\pm160\phantom{1}$ & $5.12\pm0.07$ & \citet{BrownELMi} & 0.9 \\
SDSS~J174140.49+652638.7  & 18.4 & $9790\pm240$  & $5.19\pm0.06$ & \citet{BrownELMiii} & 0.3 \\
SDSS~J210308.80-002748.89 & 18.5 & $9788\pm\phantom{1}59$   & $5.72\pm0.07$ & \citet{KilicELMiv} & 0.2 \\
SDSS~J211921.96-001825.8  & 20.0 & $10360\pm230\phantom{1}$ & $5.36\pm0.07$ & \citet{BrownELMi} & 0.6
\enddata
\end{deluxetable*}
% ==================================================================== %

Our newfound pulsator J1840 occupies a new, cooler region of the DAV instability strip, which may be an extension of the C/O-core region. The highest-amplitude period observed ($> 4400$~s) is the longest, to date, of any period observed in a DAV \citep{Mukadam06}. This makes sense qualitatively, as we would expect the periods of pulsation modes to roughly scale with the dynamical timescale for the whole star, $P \propto \rho^{-1/2}$. \citet{Ostensen10} find longer pulsation periods in lower temperature and surface gravity subdwarf B stars. A similar trend would explain the relatively long pulsation period of J1840. The period observed is a factor of 4 longer than the models of \citet{Steinfadt10} predict for an $\ell=1$, $k=10$ mode of a 0.17 \msun\ ELM WD; in the context of these models, this mode likely has an extremely high radial overtone, higher than that of any dominant mode observed in a normal-mass DAV.

We note, however, that the driving mechanism for pulsation is largely unknown for this star. It is natural to assume that the same mechanism of convective driving that operates in the C/O-core DAVs \citep{Brickhill91,Wu98,Goldreich99} is also responsible for the pulsations of low-mass WDs. This mechanism is based on the assumption that the convective turnover timescale for a fluid element, $t_{\rm to}$, is much smaller than the oscillation periods, $P_i$. For the $\log g \sim 8$ DAVs, $t_{\rm to}\sim 0.1$--1~sec and $P_i \sim 100$--1000~sec, so $t_{\rm to} \ll P_i$ is satisfied.

To estimate how this timescale scales with $g$ we note that $t_{\rm to} \sim l/v$, where $l$ is the mixing length and $v$ is the velocity of the convective fluid elements. Employing a simplified version of mixing length theory (ignoring radiative losses) we find that $v \propto (g F/\rho)^{1/3}$, where $F$ is the stellar flux and $\rho$ is the mass density at the base of the star's surface convection zone \citep{Mihalas78}. Taking the mixing length as a factor of order one times a pressure scale height, we find $l \sim c_s^2/g \propto T/g$. Putting these results together we find $t_{\rm to} \propto T (\rho/F)^{1/3} g^{-4/3} \sim g^{-4/3}$.  Thus, a $\log g \sim 6$ object would have $t_{\rm to} \sim 50$--500~sec. In order for convective driving to operate, $P_i \gg t_{\rm to}$ is required. Perhaps this is at least a partial explanation for the very long period ($> 4000$~s) seen in this pulsator. It may also set a lower limit to periods in this DAV of $\sim 500$--1000~s, although more detailed models will be needed to confirm this.  Another potential source of driving is the $\epsilon$ mechanism, i.e., driving due to the modulation of H burning in the envelope. We will address this additional mechanism after more extensive observations.

We look forward to a coordinated effort for extensive follow-up observations and analysis of this exciting new pulsating WD, and to the discovery of additional pulsating ELM WDs in order to better understand this new (or extended) DAV instability strip.

\acknowledgments

J.J.H., M.H.M. and D.E.W. gratefully acknowledge the support of the NSF under grant AST-0909107 and the Norman Hackerman Advanced Research Program under grant 003658-0252-2009. The authors are grateful to the McDonald Observatory support staff, F. Mullally for developing much of the data analysis pipeline used here, and to J. Pelletier of the UT Freshman Research Initiative for some of the analysis used in Table~\ref{tab:null}.

\end{document}